\documentclass[runningheads]{llncs}

\input{psfig.sty}
\usepackage{subfigure}
\usepackage{graphicx}
\usepackage{color}

\begin{document}

\pagestyle{headings}

\mainmatter
\title{Diophantine Networks}
\author{C. Bedogne', A.P. Masucci, G.J. Rodgers}
\institute{%
Department of Mathematical Sciences, Brunel University, Uxbridge,
Middlesex, UB8 3PH, United Kingdom}%

 \maketitle
\date{\today}
\begin{abstract}

We introduce a new class of deterministic networks by associating
networks with Diophantine equations, thus relating network topology
to algebraic properties. The network is formed by representing
integers as vertices and by drawing cliques between $M$ vertices
every time that $M$ distinct integers satisfy the equation. We
analyse the network generated by the Pythagorean equation
$x^{2}+y^{2}= z^{2}$ showing that its degree distribution is well
approximated by a power law with exponential cut-off. We also show
that the properties of this network differ considerably from the
features of scale-free networks generated through preferential
attachment. Remarkably we also recover a power law for the
clustering coefficient.

We then study the network associated with the equation
$x^{2}+y^{2}= z$ showing that the degree distribution is consistent with
a power-law for several decades of values of $k$ and that, after having reached a
minimum, the distribution begins rising again. The power law exponent, in this case, is given by $\gamma\sim4.5$
We then analyse clustering and ageing and compare our results to the ones obtained in the
Pythagorean case.

\end{abstract}

\section{Introduction}

The study of complex networks has recently attracted much attention
in the physics community and currently represents an important area
of multi-disciplinary research \cite{1}. Network theory studies the
properties of real world systems and phenomena displaying a web-like
structure, usually represented mathematically as graphs. The
individuals of the system under consideration are then symbolised as
vertices and their relations or interactions as edges. For instance
the World Wide Web is a network of web pages connected by
hyperlinks; the cell can be described as a network of chemicals
connected by chemical reactions; human language can be viewed as a
web where words are linked if they appear adjacent or one word apart
in sentences \cite{1,2}.

We recall that a network $G$  is defined formally as a couple
$(V,E)$, where $V$ is the set of vertices and $E$ is a set of
couples of vertices, called edges. If $E$ is an ordered set, the
network is said to be "directed".

A network can be specified by its  adjacency matrix $A=\{a_{ij}\}$,
where $a_{ij}=1$ if vertex $i$ is connected to vertex $j$ and
$a_{ij}=0$ otherwise. The degree $k_i$ of  vertex $i$ is then
defined as the number of nearest neighbours of vertex $i$, that is:
$k_i=\sum_j a_{ij}$. The neighborhood $\Omega_i$ of a vertex $i$ is
the set of all vertices connected to $i$. Notice that, in contrast
with the terminology usually introduced in analysis and general
topology, $\Omega_i$ does not contain $i$ as one of its elements.
The degree function $k_i$ then "counts" the number of vertices
belonging to each neighborhood and gives thus a first
characterization of a network's topology. In many cases a more
complete characterization of network topology is given by the
weighted adjacency matrix $W=\{w_{ij}\}$, where the numbers $w_{ij}$
represent the weight of the connection between vertex $i$ and vertex
$j$. This number can be defined in many ways. In our case we will
define it as the total number of edges connecting vertex $i$ and
vertex $j$. We then define the strength $s_i$ of vertex $i$ as the
total number of edges belonging to vertex $i$, that is
$s_i=\sum_jw_{ij}$. When multiple edges are prevalent a weighted
analysis is obviously worthwhile. More refined characterizations of
network topology also take account of clustering \cite{1} and of the
average distance between two vertices \cite{1}.

Complex networks were originally studied in the context of random
graph theory \cite{1,4} where it has been shown that connecting
the vertices at random results in  exponential degree
distributions. The structure of a random graph is thus uniform,
with most vertices having approximately the same degree and only
few vertices of high degree.

Empirical results indicate however that several real-world
networks display instead power-law degree distributions \cite{1}
$p(k)= ck^{-\gamma}$, where $c$ and $\gamma$ are constants.
Because of the self-similarity properties of the power law, these
networks are called scale-free. It is very remarkable that both
the empirical and the analytical work usually recovers scale-free
networks with power law exponents ranging between $2$ and $3$.

Since power laws decay much slower than exponentials, scale-free
topologies are not uniform and display many highly connected
vertices, called hubs. Hubs work as shortcuts between apparently
distant environments and thus have  a crucial role in many
phenomena, such as in the spreading of infective agents through a
network \cite{5} or resilience of the network  to external attacks
\cite{6}.

Scale-free topologies are normally interpreted as a consequence of
network self-organization, resulting in properties of the system
\emph{as a whole} which (in contrast with classical physics) cannot
be inferred only from properties of its \emph{parts} (that is from
the interactions between the individual components considered in
isolation).

Thus one of the major problems in complex network theory has been to
identify  the range of mechanisms by which a network can
self-organise into a scale-free state \cite{7}.

It has been shown analytically \cite{1,8,9,10} that if a network
evolves by addition of new vertices at a constant rate and if the
newly introduced vertices connect \emph{preferentially} (and
\emph{linearly}\cite{8}) to highly connected vertices, then the
network displays a scale-free topology.

Preferential attachment, however, is not always a natural
hypothesis \cite{6,10,11}. For example, in many situations it is
not realistic to assume that a node has the complete information
about the degree distribution it would need in order to know where
to attach preferentially. From the other hand this information is
just what is needed to normalise properly \cite{34} the attachment
probabilities as defined in most of the analytic models introduced
in literature so far \cite{1,7}). Alternative mechanisms
generating scale-free topologies have then been introduced, as for
example the static model \cite{13}, the varying fitness model
\cite{11,12,R1,R2,K1} and random walk models incorporating a
copying mechanism \cite{34}.

In addition to random networks, some interesting examples of
\emph{deterministic} networks have also been recently studied. A
remarkable example are the Apollonian networks discussed in
\cite{14,15,16,17}, inspired by the ancient problem of finding a
space filling packing of spheres. Apollonian networks turn out to
be simultaneously scale-free, small world, Euclidean,
space-filling and matching graphs. In this case the scale-free
topology, however, turns out to be implicitly related to a
preferential attachment mechanism \cite{16}.

It is important to recall that deterministic networks are, at least
in principle, completely controllable and can thus  be very useful
in the design of artificial networks such as communication or
economic networks, where (depending on the particular problem under
consideration) the topology will usually have to satisfy several
spacial and temporal constraints.

The integer networks introduced in \cite{18} are another
interesting example of  deterministic networks where vertices
represent positive integers and edges are drawn whenever there is
a divisibility relation between them. These networks are directed
and their topology crucially depends on the particular set of
vertices under consideration. For instance, when the  vertices
range in the set of the prime numbers an infinite star network
centered in $1$ is obtained. Considering instead the set of
composite numbers, the degree distribution is the sum of an
in-degree distribution and an out-degree distribution, and
numerical simulations indicate that the latter is well
approximated by a power-law with exponent $2$.

Thus our attention is drawn to abstract networks
whose vertices are mathematical objects and whose edges symbolise
mathematical relations. Examples of this kind obviously abound. A
straightforward example is the set of the infinitely
differentiable functions defined in the unit interval. We identify
functions (through an equivalence relation) when they differ by a
constant and naturally define a network structure on the set of
equivalence classes by connecting vertex $f$ (for the sake of
notational economy here we denote the equivalence classes by their
representants) to vertex $g$ if $g$ is the derivative of $f$. Each
non-constant vertex $f$ has thus a descendent (its derivative
$f'$) and every vertex has an antecedent (its integral $F(x)=
\int_{0}^{x} f(t) dt$). We thus have defined a directed network
consisting  of infinite disconnected components. The degree of
each vertex can only take the values $1$ or $2$ and the components
can be described as infinite branches, finite branches (such is
the case when one of the vertices of the branch is a polynomial),
cycles (such as the one generated by $sin x$) and a loop
(generated by $exp(x)$).

Another way to associate a network to mathematical objects is by
associating networks to equations. Consider any equation $F(x_{1},
x_{2},..., x_{n})=0$, where the variables $x_{1},x_{2},..., x_{n}$
range in a given set $S$. We can then naturally associate a network
to the equation by representing the points of $S$ as vertices and by
forming an n-clique (that is a completely connected subgraph) each
time that the $n$ elements of $S$ satisfy the equation. We will then
say that the network thus defined is \emph{generated} by the
equation $F(x_{1}, x_{2},..., x_{n})=0$.

Cliques have been widely studied in the last few years \cite{19} and
often provide interesting insights into network structure and
function. For instance cliques naturally represent clusters,
communities and groups in social networks. Cliques are also very
important in theoretical biology when modelling protein-protein
interaction networks \cite{20} and gene regulatory networks
\cite{21}.

In this paper we will study the degree distribution of networks
generated by Diophantine equations, that is polynomial equations
whose variables range in the set of integer numbers (originally
introduced by Diophantus of Alexandria), attempting thus to establish a
connection between age old unsolved problems in number theory and
a very actual and practical area of Science.

In particular, in Model A we study the \textit{finite} networks
generated by the Pythagorean equation $x^{2}+y^{2}=z^{2}$ when the
variables range in the set $\{1,2,..., N\}$ recovering (at least up
to $N=10^{5}$) a degree distribution that  is well approximated by a
power law with exponential cut-off. An ageing analysis reveals then
that the hubs of the network, which are the nodes with the highest
degree, are not the oldest vertices of the network  as we would
expect for a network grown via preferential attachment. In the
Pythagorean case hubs form instead in vertices whose age is between
old and middle age. After a while a freezing effect takes place and
the hubs stop attracting new connections.Younger vertices then take
over and become the new hubs in the network. This situation looks
more akin to real social networks than to the $rich$ $get$ $richer$
networks. This idea has already been suggested in a recent work
\cite{22}, in which rules are added to a stochastic network. The
basic suggestion is that people in social networks agree and form
hubs in response to an exact need, a deterministic rule, and not
merely to an unbiased attraction to important people. The clustering
coefficient analysis also displays interesting properties. In
particular in Pythagorean networks $c(k)$ has a power law behaviour
that can be predicted analytically.

In Model B we introduce the finite networks generated by the Diophantine
equation $x^{2}+y^{2}=z$ in the set $\{1,2,..., N\}$. We analyse the degree distribution and
clustering coefficient comparing the results to those obtained in Model A.
In particular we  find that, although the network's topology turns out to be
very different than the one obtained starting from the Pythagorean equation, a power-law can
still be recovered for several decades of values of $k$.

\section{Model A}

We consider  the network generated by Diophantine equations of the
form
\begin{equation}
x^{n}+y^{n}=z^{n}
\end{equation}
where the variables $x$, $y$, $z$ and the exponent $n$ are
\emph{positive} integers. We study how the network generated by Eq.1
evolves by introducing a new vertex at each time-step, following the
natural order of the positive integers. Thus at time-step $t=1$ the
vertex $1$ is introduced and no edge forms. At each subsequent time
step another integer is introduced and $3$-cliques form whenever
triplets of integers satisfy Eq.1.

We then study the  \emph{finite }networks obtained by limiting the
variables to range within sets of the kind $\{1,2,...,N\}$. The
network will thus consist of $N$ vertices (associated with the first
$N$ positive integers) and a number of edges belonging to 3-cliques.
Notice first of all that at each time step the degree of every
vertex is expressed by an even number. This fact alone has some
interesting consequences in terms of network topology. For instance,
it follows from a well-known theorem of graph theory that each
connected component of the finite networks we just defined has an
Eulerian circuit, that is a circuit joining all edges and traversing
each edge only once. Notice also that, according to Fermat's Last
Theorem, proven by Wiles in 1994, Eq.1 has no solutions for $n\geq3$
and the corresponding networks are thus empty. We therefore need to
consider only the cases $n=1$ and $n=2$.

\subsection{The linear case}
We consider first the linear case
\begin{equation}\label{l0}
x+y=z
\end{equation}
with $z< N$.

In this case it is a simple matter to derive  the degree
distribution analytically. Since multiple links are prevalent we
performed a weighted analysis. We recall that the weight of the
links is defined as the number of times two vertices are connected
and the strength of a vertex as the total number of vertices it is
connected to, having counted the repeated links. Since we allow
loops, derived by a solution of Eq.\ref{l0} of the type $x+x=2x$,
every vertex is connected to all other vertices if $2x<N$, and to
all other vertices but itself if $2x\geq N$, so that $k(x)=N-1$ if
$x<\frac{N}{2}$ and $k(x)=N-2$ if $x\geq \frac{N}{2}$. If $N$ is
even we have
$P(k)=\frac{N-2}{2(N-1)}\delta(k-N+1)+\frac{N}{2(N-1)}\delta(k-N+2)$.
If $N$ is odd,
$P(k)=\frac{1}{2}\delta(k-N+1)+\frac{1}{2}\delta(k-N+2)$.

To calculate the strength of the vertices we have to consider that,
if $n<\frac{N}{2}$, the vertex $n$  will assume $2(N-1-n)+1$ links
(note that in this case the clique $(n,n,2n)$ will give 3 new links
to vertex $n$ ), forming the following $N-1-n$ cliques:
$(n,1,n+1),(n,2,n+2)..., (n,N-1-n ,N-1)$, plus $n-1$ links with the
following cliques: $
(1,n-1,n),(2,n-2,n),...,(\frac{n}{2},\frac{n}{2},n)$ if $n$ is even
and $ (1,n-1,n),(2,n-2,n),...,(\frac{n+1}{2}-1,\frac{n+1}{2},n)$ if
$n$ is odd. If $n\geq\frac{N}{2}$ the vertex $n$ will assume
$2(N-1-n)$ links forming the following $N-1-n$ cliques:
$(n,1,n+1),(n,2,n+2)..., (n,N-1-n ,N-1)$ plus $n-1$ links with the
following cliques: $
(1,n-1,n),(2,n-2,n),...,(\frac{n}{2},\frac{n}{2},n)$ if $n$ is even
and $(1,n-1,n),(2,n-2,n),...,(\frac{n+1}{2}-1,\frac{n+1}{2},n)$ if
$n$ is odd.  It follows that the strength of a vertex $n$ is given
by $s(n)=2N-n-2$ if $n< \frac{N}{2}$ and by $s(n)=2N-n-3$ if $n\geq
\frac{N}{2}$. Thus linearity of Eq.2 results  in a uniform
probability distribution for the strength: $P(s)=\frac{1}{N}$ if
$N-3\leq s\leq 2N-3$ and $P(s)=0$ otherwise.

Notice at last that the degree distribution, in the thermodynamic
limit, becomes trivial as each vertex's degree diverges when
$N\longrightarrow\infty$.

\subsection{The Pythagorean case}

We now study  the network generated by the Pythagorean equation
\begin{equation}\label{P1}
x^{2}+y^{2}=z^{2}
\end{equation}
Following standard mathematical terminology, we will say that
$(x,y,z)$ is a Pythagorean triplet whenever $x$,$y$ and $z$ satisfy
Eq.3. The numbers $x$, $y$ and $z$ are usually called the "legs" of
the triplet. When all legs are positive we say that $(x,y,z)$ is a
positive Pythagorean triplet. A Pythagorean triplet is called
"primitive" whenever $x$, $y$ and $z$ are relatively prime. It has
been known since antiquity that every Pythagorean triplet is either
of the form $(\lambda (a^{2}-b^{2}), 2\lambda ab, \lambda (a^{2}+
b^{2})$ or of the form $(2\lambda ab,\lambda (a^{2}-b^{2}), \lambda
(a^{2}+ b^{2}) )$. It is also well known that every positive integer
$x>2$ belongs to at least one Pythagorean triplet \cite{25} and that
there are infinite primitive Pythagorean triplets but, despite the
apparent simplicity of the problem, the topological distribution of
the set of all Pythagorean triplets is still largely unknown
\cite{22,23,24,25,26,27}.

It is known for instance that the number $A(N)$ of Pythagorean
triplets with legs smaller or equal to a given positive integer $N$
can be expressed by $A(N)=\frac{4}{\pi}N logN+BN+E(N)$ where $B$ is
an explicitly given constant and $E(N)$ is a remainder function. It
is also known that $E(N)= O (N^{\frac{1}{2}})$ and there is a large
literature aimed at giving a better estimate for the remainder
\cite{22,23,24,25,26,27}. It was proved in \cite{26} that $E(N)=
O(N^{\frac{1}{2}} exp(-c(log N)^{\frac{3}{5}}(log log
N)^{\frac{1}{5}})$ with some $c>0$. Notice however that, in order to
investigate the Pythagorean network just introduced analytically and
derive its degree distribution we would clearly need to know much
more, that is the exact number $P(x)$ of Pythagorean triplets of
which $x$ is a member. As far as we are aware this is still an open
problem in number theory, probably deeply connected with the
factorization problem. Indeed it is easy to conjecture that the
properties of $P(x)$ don't depend only on the algebraic form of Eq.3
but also on the structure of the number $x$ itself in terms of its
factorization as product of prime numbers. For instance it is known
that if $w(x)$ is the number of distinct prime factors of an integer
number $x>1$, then the number $P(x)$ of Pythagorean triplets of
which $x$ is a member is given by $P(x)= 2^{w(x)}$ if all prime
factors of $x$ satisfy $p\equiv 1 mod 4 $,$P(x)= 2^{w(x)-1}$ if $x$
is odd and some prime factor of $x$ satisfies $p\equiv 3 mod 4 $,
$P(x)= 2^{w(x)-1}$ if $n\equiv 0 mod 4 $ and $P(x)=0$ otherwise. The
problem of determining the distribution of Pythagorean triplets is
thus reducible, at least within the set of \emph{primitive} triples,
to the problem of counting the prime factors of each leg, which
strongly suggests that there is little hope to derive the degree
distribution analytically at present. The model, however, can be
described numerically, at least in the finite case. We will then
consider the networks generated by allowing the variables of Eq.3 to
range only in the first $N$ \emph{positive} integers.

\subsubsection{Numerical analysis}
\begin{figure}[!ht]\center
         \includegraphics[width=0.48\textwidth]{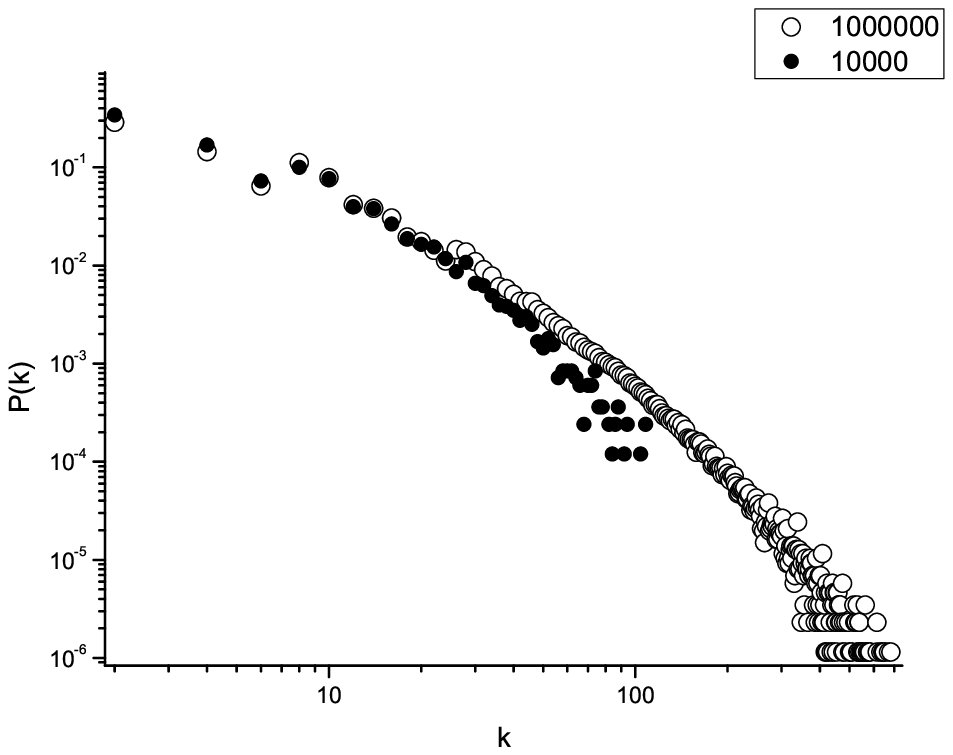}
         \includegraphics[width=0.48\textwidth]{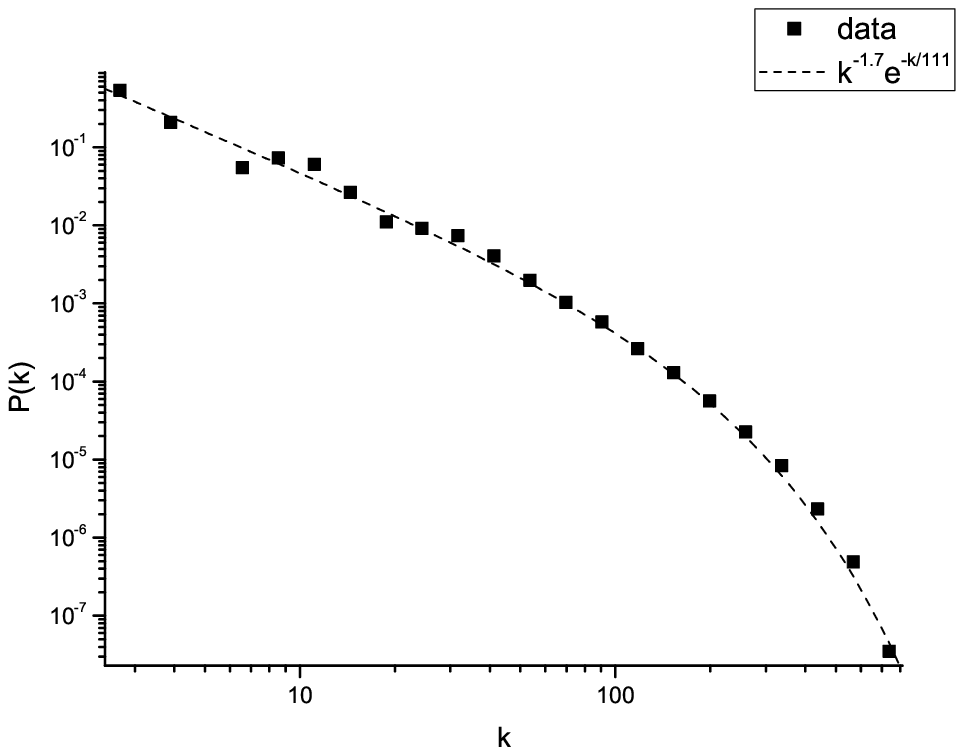}
\caption{\label{p1} In the left panel we show the degree
distribution for the Pythagorean network for $z<10^4,10^6$.
In the right panel we perform a logarithmic binning on the degree
distribution for $z<10^6$ to illustrate the power law behaviour. The fitting curve is the power law with exponential cut-off in Eq.\ref{n10}.}
\end{figure}

We performed numerical simulations of the network growth with the
deterministic attachment rule given in Eq.3. We show the results for
the degree distribution in the left panel of Fig.\ref{p1} for
$z<10^4,10^6$. What is  first evident from the log-log plot is that,
changing the scale of the network size, the degree distribution
preserve its shape. Then, with increasing values of $z$, a wide
region emerges where the probability distribution looks like to
follow a power law.  On the right panel of Fig.\ref{p1} we show the
empirical distribution for $z<10^6$ after a logarithmic binning to
understand better the function behind the data. We find that a power
law with an exponential cut-off,
\begin{equation}\label{n10}
P(k)\propto e^{-\frac{k}{111}}k^{-1.7},
\end{equation}

 approximates very well the
empirical data.

Considering that complex growing networks usually display pure
power law distributions only in the thermodynamic limit \cite{22}, we find
the finite results expressed graphically in Fig.1 particularly
remarkable.

\begin{figure}[!htbp]\center
         \includegraphics[width=0.48\textwidth]{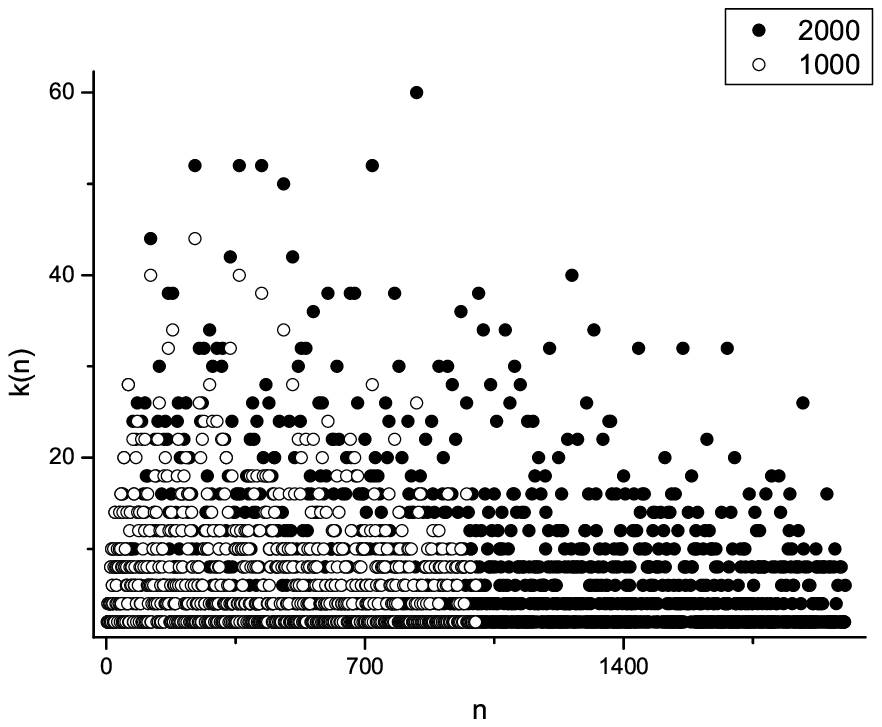}
         \includegraphics[width=0.48\textwidth]{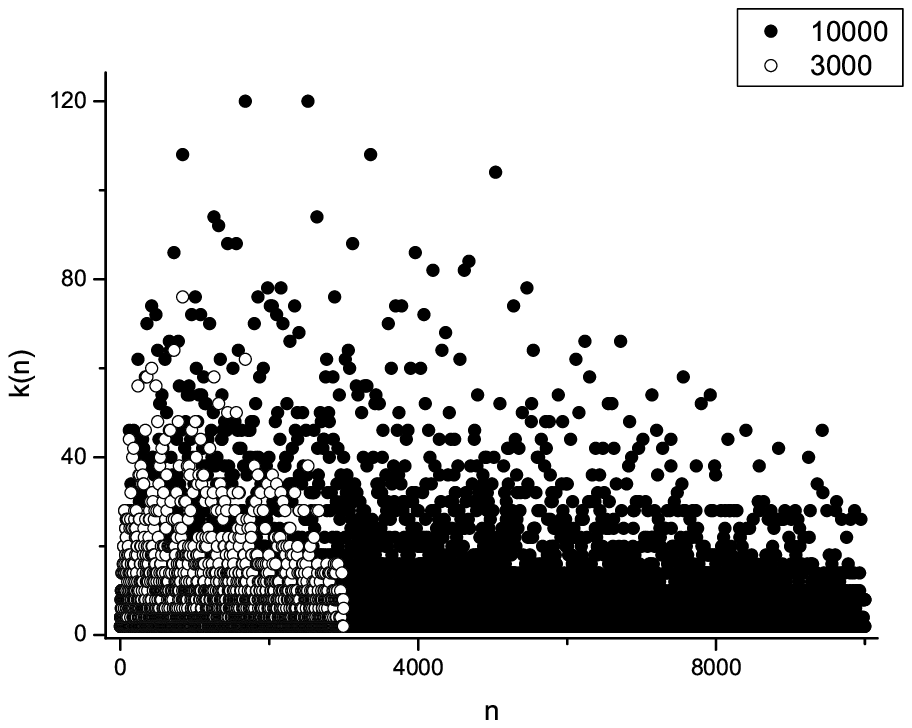}

\caption{\label{p2} Representation of the aging of the hubs for
the Pythagorean network for $z<1000,2000$ on the left and for
$z<3000,10000$ on the right. $k$ is the degree of the vertices,
while $n$ is their label. Small values of $n$ correspond to old
vertices.}
\end{figure}

To better understand how those structures form, we performed an
$ageing$ analysis of the network. In Fig.\ref{p2} we show the
degree of the vertices versus their labels for different values of
$z$. Small labels indicate old vertices. Peak values of the degree
$k$ represent the major hubs of the network. For stochastic
preferential growing networks \cite{28}, the high degree vertices
are always the old vertices of the network. In the Pythagorean
network this result does not hold. From Fig.\ref{p2} we can see
that after a while the old vertices' degree freezes and hubs form
between young and middle-aged vertices.

\begin{figure*}
\includegraphics[width=9cm]{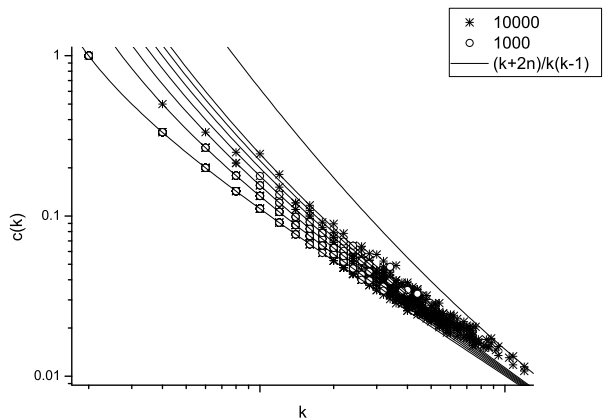}
\caption{\label{p3} Clustering coefficient for the Phytagorean
network measured in simulations with $z<1000,10000$. The lines
represent Eq.\ref{n1} for different values of $n$. In particular the
greatest value of $n$ allowed for $z<10000$ is $n=23$. }
\end{figure*}

Since the Pythagorean network is made up of cliques it is
interesting to look at its clustering coefficient.The clustering
coefficient \cite{1} is defined as
\begin{equation}\label{n0}
c(k_i)=\frac{2e_i}{k_i(k_i-1)},
\end{equation}
where $k_i$ is the degree of vertex $i$ and $e_i$ is the number of
nearest neighbors of vertex $i$ that are connected to each other. We
show in Fig.\ref{p3} our measures of $c(k)$ on a Pythagorean network
for different values of $z$. The regularity of the distribution is
striking and due to the high regularity of the network. All vertices
with $k=2$ form a triangle with their nearest neighbors, so that
they have $c(k)=1$. The higher a vertex degree, the more improbable
it is that its nearest neighbors are fully connected. Since for the
inner geometry of the network $e_i\geq \frac{k_i}{2}$, it follows
that $c(k_i)\geq \frac{1}{k_i-1}$ and, obviously, $c(k_i)\leq 1$.
All the values of $c(k_i)$ lie on well-determined levels defined by
\begin{equation}\label{n1}
c(k_i)=\frac{k_i+2n}{k_i(k_i-1)},
\end{equation}
where $n=0,1,2,...,n_i$ and $n_i=\frac{k_i(k_i-2)}{2}$. To
demonstrate Eq.\ref{n1} is a simple graphical exercise.
Eq.\ref{n1} is compared to the numerical simulation in
Fig.\ref{p3}. For each $k_i$, the different values of $c(k_i)$ are
equally spaced at a distance of $\Delta
(c(k_i))=\frac{2}{k_i(k_i-1)}$.

Notice at last that, in contrast with the linear case, the
network's behaviour in the thermodynamic limit is now highly not
trivial and probably cannot be described in detail before some
open questions in number theory will be eventually answered. The
numerical work we performed may however suggest some interesting
conjectures. We think that the most relevant feature of our model
is perhaps the ageing analysis showing that, when the network
grows, the hubs are constantly rejuvenated, in sharp contrast with
the behaviour of scale-free networks generated through the
preferential attachment mechanism.

\section{Model B}

We now study the network generated by the deterministic rule
expressed by the equation
\begin{equation}\label{he1}
x^{2}+y^{2}=z
\end{equation}
This network is thus deeply connected with the classical Euler and
Fermat Problem of finding all integers that can be expressed as the
sum of two squares. In a sense this case is intermediate between the
linear and the Pythagorean case. Notice, first of all, that in the
thermodynamic limit Eq.\ref{he1} generates a trivial network where
each vertex's degree goes to infinity. Indeed given any positive
integer $x$, for any arbitrary integer $y$ (when
$N\longrightarrow\infty$) we can always find an integer $z$
satisfying Eq.\ref{he1}. Thus we only need to consider the finite
networks specified by the inequality $z<N$. The network attachment
mechanism, as in Model A, will depend on the generating equation in
the sense that every triple of integer numbers obeying Eq.\ref{he1}
forms a clique in the network.

\begin{figure}[!htbp]\center
         \includegraphics[width=0.48\textwidth]{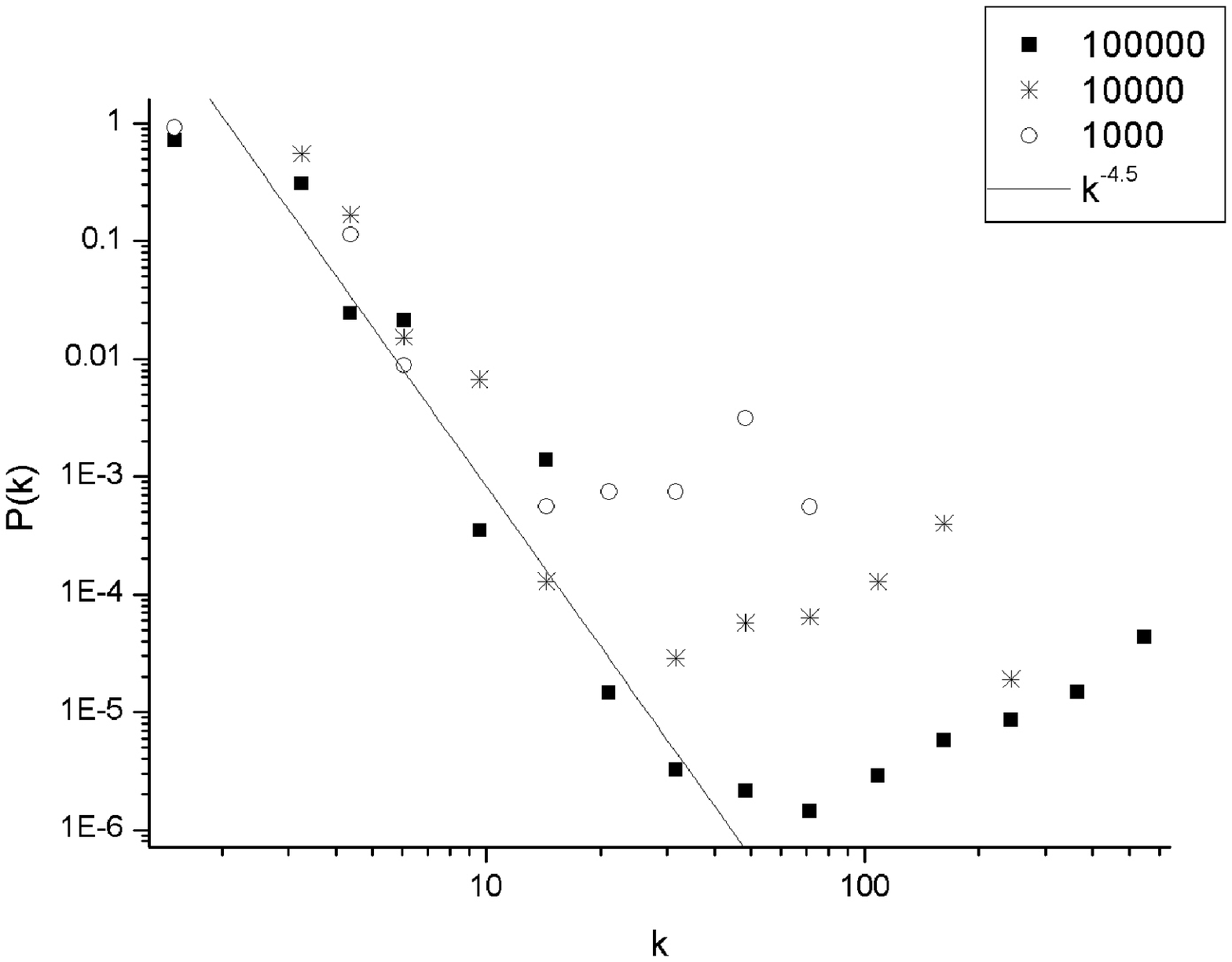}
         \includegraphics[width=0.48\textwidth]{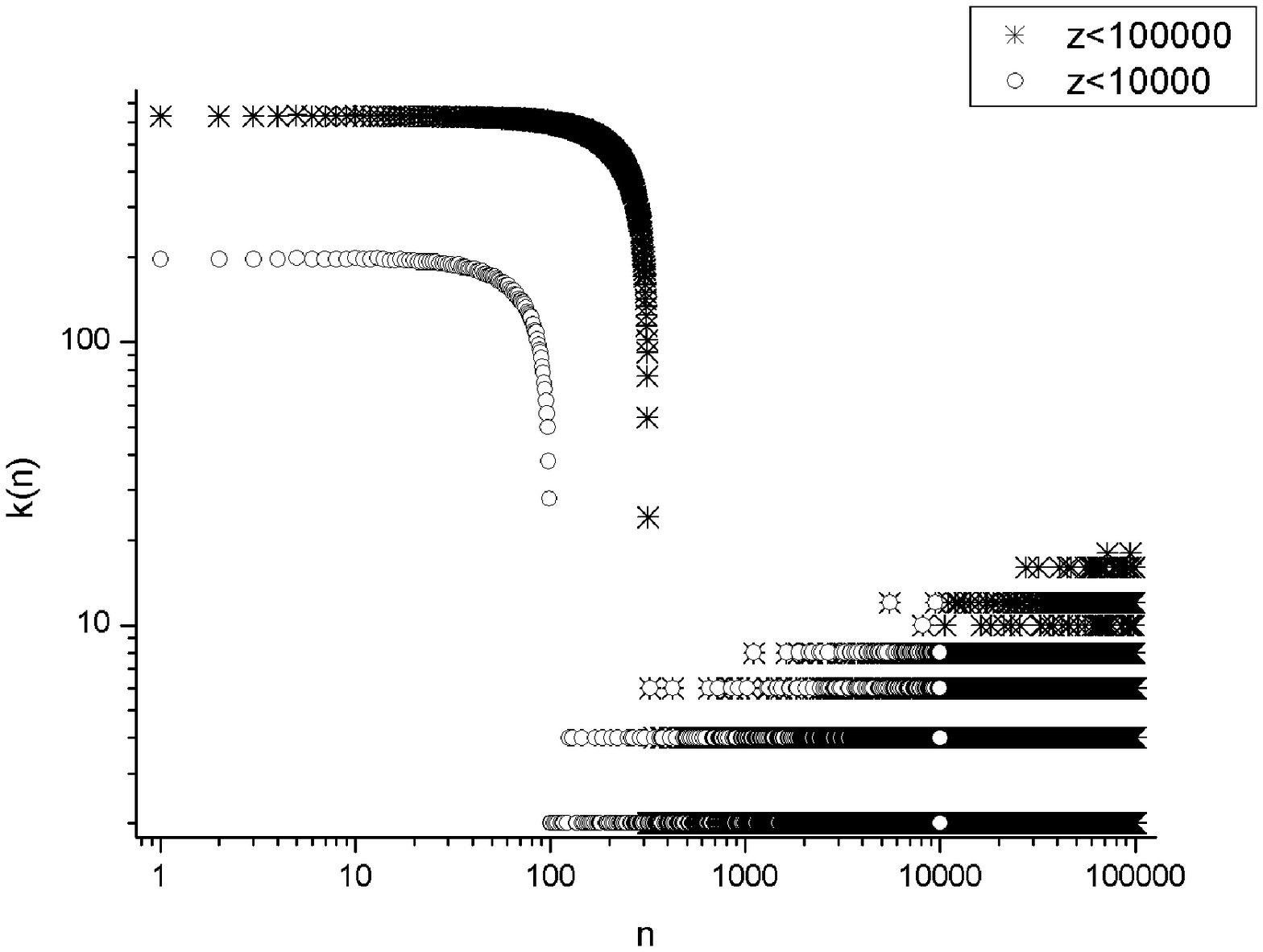}

\caption{\label{h1} Left panel: degree distribution for the
deterministic network generated by the rule in Eq.\ref{he1} for
$z<1000, 10000, 100000$. Right panel: representation of the aging of
the hubs for the same network for $z<10000,100000$. $k$ is the
degree of the vertices, while $n$ is their label. Small values of
$n$ correspond to old vertices.}
\end{figure}

The resulting network is very well connected since every couple of
numbers at a certain time step (when $N$ is sufficiently large)
forms a clique satisfying Eq.\ref{he1}. Nevertheless interesting
topological properties emerge. In the left panel of Fig.\ref{h1} we
show the degree distribution for this network. The shape of the
distribution is well approximated by a power law with exponent
$-4.5$ for several decades of values of $k$. Very interestingly,
after falling the distribution begins to rise again. On the right
panel of the same figure we show the degree of the vertices against
their label number $n$. This number can be considered as the birth
time of the vertex so that we can observe the evolution of the hubs
of the network with ageing. The degree is nearly constant and very
high for small values of $n$ and it drops rapidly to small values
for $n \sim \sqrt{N}$ (this comes from Eq.7, when $z\sim N$ then
$x,y\sim \sqrt{N}$).

Notice that the power law exponent is considerably lower (at least
from the point of view of network theory where power law exponents
usually range between $-3$ and $-2$) than the one we recovered in
the Pythagorean case. We interpret this weakening as the result of
the upper thresholds we imposed to the network which, in this
case, have a much stronger effect on the degree distribution than
in the Pythagorean case.

\begin{figure}[!htbp]\center
         \includegraphics[width=0.48\textwidth]{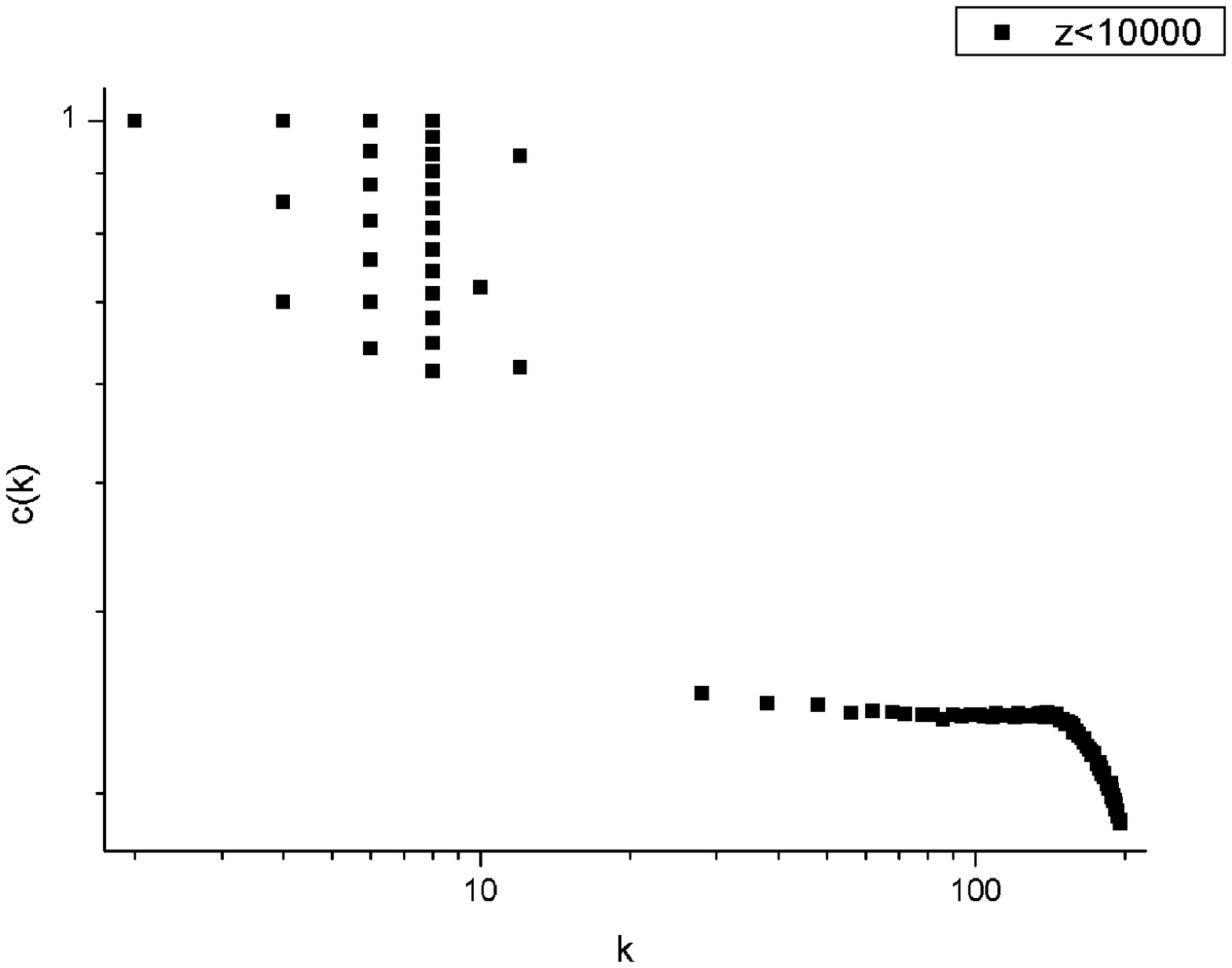}
         \includegraphics[width=0.48\textwidth]{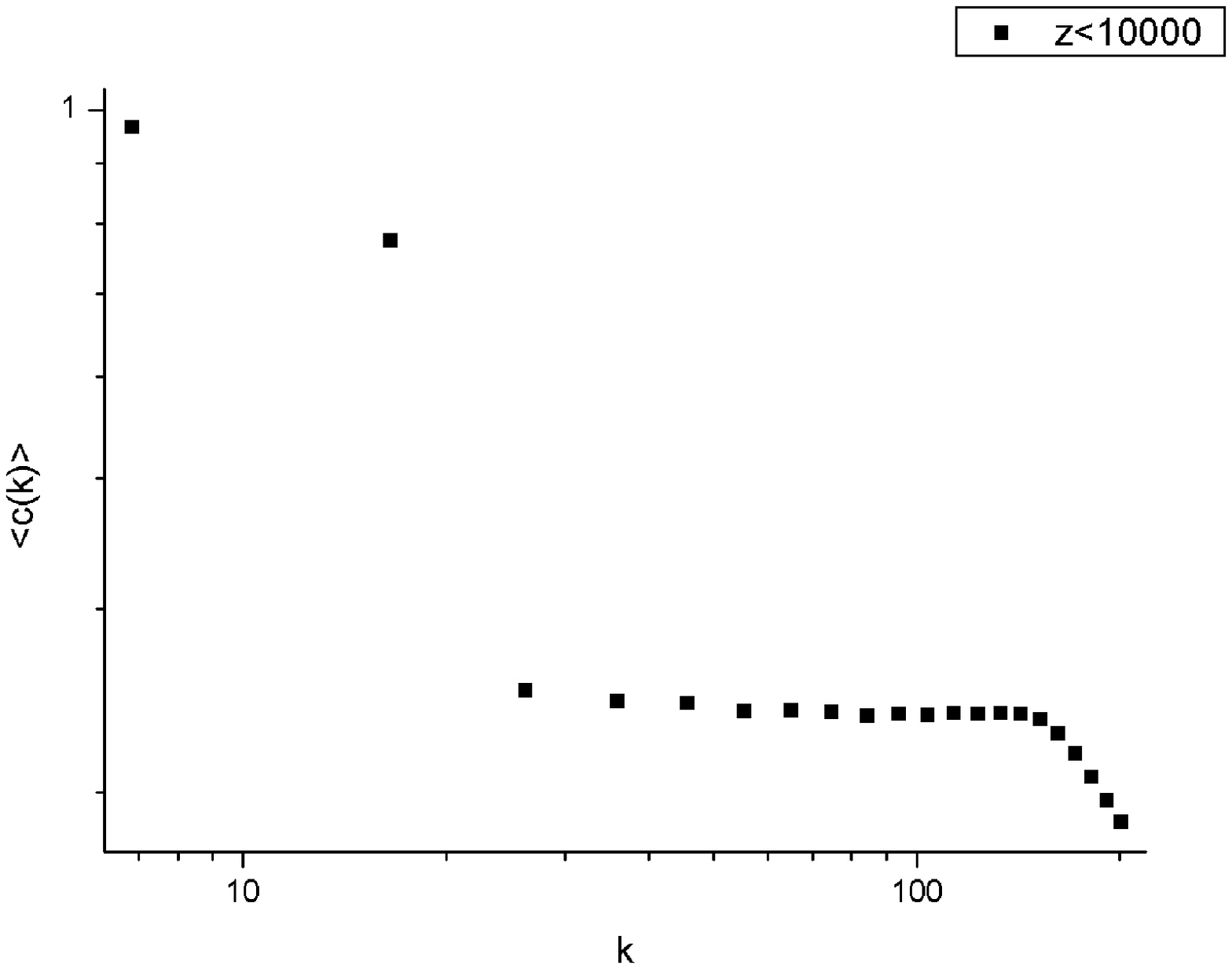}

\caption{\label{h2} Left panel: clustering coefficient $c(k)$
against the degree $k$ for the deterministic network generated by
the rule in Eq.\ref{he1} for $z<10000$. Right panel: average
clustering coefficient $c(k)$ against the degree $k$ for the
deterministic network generated by the rule in Eq.\ref{he1} for
$z<10000$.}
\end{figure}

In Fig.\ref{h2} we show the clustering analysis for the network. On
the left panel we show the clustering coefficient $c(k)$, defined in
Eq.\ref{n0}, against the degree, while on the right panel we show
the average clustering coefficient $<c(k)>$ against the degree. We
can observe that it is very high for small values of $k$, where the
degree distribution is a power law, then it is nearly constant for
several decades and finally it falls rapidly to zero for high values
of the degree. This behaviour has to do with the clique structure of
the network. $c(k)=1$ as soon as isolated cliques form, that is for
$k=2$, then its value falls as new connections are added and the
degree increases.

\

\textbf{Conclusions}

We have introduced the networks generated by Diophantine equations
focusing on two classical problems of number theory widely studied
since antiquity. We investigated first the Pythagorean problem of
finding all triples of positive integers that can be represented
geometrically as legs of a  right-wristed triangle. We studied then
the (closely related) problem of determining the set of positive
integers that can be represented as the sum of two squares. These
two questions, despite their apparent simplicity, have not yet been
answered analytically in their full generality.

We noticed that these classical problems can be naturally translated
into the language of network theory and studied them numerically in
their  finite versions (that is by not allowing integers to be
greater than a fixed upper threshold). It is then easy to see that
the equations considered generate clique structured networks, in an
interesting analogy with many relevant real world networks such as
social \cite{31} and biological networks \cite{20,21}.

In the Pythagorean case, where the network evolves through the
deterministic attachment mechanism specified by Eq.3, we recovered a
degree distribution consistent with a power law with exponential
cut-off, in accordance with the behaviour displayed by many real
world random networks \cite{1}.

We then performed an ageing analysis and showed that, in sharp
contrast with stochastic preferential growing networks, hubs form
between young and middle aged vertices (the degree of each vertex,
after a while, freezes).

The power-law, in this case, is clearly not a by-product of any
preferential attachment or fitness mechanism, but a consequence of
the algebraic structure of Eq.3 itself. This shows once more that
preferential attachment is a sufficient but not a \emph{necessary}
condition to the generation of scale-free topologies. As far as we
are aware the network generated by Eq.3 is the first genuine example
in literature of a \emph{deterministic} scale-free network (at least
in some approximation)  which does not grow preferentially.

Power laws are normally regarded as a sign of complexity and we
found it particularly stimulating to detect them not only in natural
phenomena but also at the roots of number theory itself, indicating
a fascinating connection between complex systems and pure
mathematics.  We then studied the second classical problem mentioned
above studying the network generated by Eq.\ref{he1} and recovered a
degree distribution approaching a power-law with a relatively high
exponent $\gamma\sim4.5$ for several decades of values of $k$.

We believe that the study of networks generated by equations will
turn out to represent an important field of interdisciplinary
research and we hope that our discussion will be useful to
researchers in both complex network theory and pure mathematics as
a first step to recover more rigorous and general results. In
particular, in future work it will be stimulating to investigate
the relation between an equation's algebraic properties (or
analytic properties if, for instance, differential equations are
considered) and the associated network's topology.

Diophantine networks may find interesting theoretical and technological applications.
First and foremost they can be used to design deterministic toy models of complex systems, allowing to find practical ways
to build networks systematically
without having to deal with different degrees of stochasticity in
their architecture (in this sense they could play a role similar
to the one played by the Euclidean grid in other traditional contexts).
Diophantine networks  may also allow to overcome some limitations
intrinsic in the preferential attachment method, namely the fact
that the hubs are usually the oldest vertices and never "die",
leaving little room for network rejuvenation. Here instead we have
a dynamic situation where hubs are dominant, during network
growth, only for a limited period of time. There may be situations
where this feature is realistic (for instance in models of
technological developments). Another interesting application of the
theory may be found in the study of the behaviour of dynamically
evolving agents.

\

\textbf{Acknowledgments }

\

We thank the European Union Marie Curie Program (NET-ACE project,
contract number MEST-CT-2004-006724) for financial support.

\


\thebibliography{apsrev}


\bibitem{1} R.Albert, A.L.Barab\'{a}si, Rev.Mod.Phys.
\textbf{74}, 47 (2002)

\bibitem{2} A.P.Masucci, G.J.Rodgers, Phys.Rev.E \textbf{74}, 046115 (2006)

\bibitem{4}B.Bollob\'{a}s, \textit{Random Graphs} (Academic Press,
London, 1985)

\bibitem{5} R.Pastor-Satorras, A.Vespignani, \textit{Scale-Free Networks} (Wiley-VCH,
Berlin, 2002), chapter Epidemics and Immunization

\bibitem{6} R.Albert, H.Jeong, A.L.Barab\'{a}si, Nature
\textbf{406}, 378 (2000)

\bibitem{7} C.Bedogne', G.J.Rodgers, Phys.Rev.E \textbf{74}, 046115 (2006)

\bibitem{8}P.L.Krapivsky, S.Redner, F.Leyvraz,
Phys.Rev.Lett. \textbf{85}, 4629 (2000)

\bibitem{9} P.L.Krapivsky, S.Redner, Phys.Rev.E
\textbf{63}, 066123 (2001)

\bibitem{10}P.L.Krapivsky, G.J.Rodgers, S.Redner, Phys.Rev.Lett. \textbf{86}, 5401 (2002)

\bibitem{11} V.D.P.Servedio, G.Caldarelli, Phys.Rev.E
\textbf{70}, 036126 (2004)

\bibitem{12} G.Caldarelli, A.Capocci, P.De Los Rios, M.A.
Munoz, Phys.Rev.Lett. \textbf{89}, 258702 (2002)

\bibitem{R1} G.Ergun, G.J.Rodgers, Physica A \textbf{303}, 261 (2002)

\bibitem{R2} K.Austin, G.J.Rodgers, Proc.of the Int.Conf.of Comput.Science ICCS2004, 1087 (2004)


\bibitem{K1} W.Hwang, P.L.Krapivsky, S.Redner, J.Math.Biol. \textbf{44}, 375 (2002)

\bibitem{13} K.I.Goh, B.Kahng, D.Kim, Phys.Rev.Lett.
\textbf{87}, 278701 (2001)

\bibitem{14}  Z.Zhang, F.Comellas, G.Fertin, L.Rong, J.Phys.A.Math.Gen. \textbf{39}, 1811 (2006)

\bibitem{15}  Z.Zhang, R.Lili, Z.Shuiseng, Phys.Rev.E \textbf{74}, 046105 (2006)

\bibitem{16}  J.S.Andrade Jr, H.Herrmann, R.F.S.Andrade, L.R.da Silva, Phys.Rev.Lett. \textbf{94}, 018702 (2005)

\bibitem{17}  R.F.SD.Andrade, J.G.V.Miranda, Physica A \textbf{356}, 1 (2005)

\bibitem{18}  T.Zhou,B.H.Wang, P.M.Hui, K.P.Chan, Physica A \textbf{367}, 613 (2006)

\bibitem{19}  K.Tokemoto, C.Oosawa, T.Akutsu, Physica A \textbf{380}, 665 (2007)

\bibitem{20}  R.Milo \textit{et al}, Science \textbf{298}, 824 (2002)

\bibitem{21}  S.Shen-Orr, R.Milo, S.Mangan, U.Alam, Nat.Genet.\textbf{31}, 64 (2002)

\bibitem{22}  G.Ghoshal, M.E.J.Newman, arXiv:physics/0608057v2 (2007)

\bibitem{23} A.Ivic \textit{et al}, arXiv:math.NT/0410522v1 (2004)

\bibitem{24} E.K.Hinson, Fibonacci Quart. \textbf{30}, 335 (1992)

\bibitem{25} M. Benito, J.L.Varona, J. of Comput. and Appl.Math. \textbf{143}, 117 (2002)

\bibitem{26} F.Wiedijk, Formalized Mathematics \textbf{9}, 809 (2001)

\bibitem{27}M. Kuehleitner, Abh.Math.Semin.Univ.Hamb. \textbf{63}, 105 (1993)

\bibitem{28} J.Lambek, L.Moser, Pacific J.Math. \textbf{5}, 73 (1955)

\bibitem{31} J.Scotts, \textit{Social Network Analysis: A Handbook}, 2nd Ed. Newberry Park, CA: Sage (2000)

\bibitem{32}D.J.Watts, P.J.Dodds, M.E.Newman, Science \textbf{296}, 1302 (2002)

\bibitem{33}G.Palla, I.Der\'{e}ny, I.Farkas, T.Vicsek, Nature \textbf{435}, 207 (2005)

\bibitem{34}T.S.Evans, arXiv:0711.0603v1 (2007)
\end{document}